\documentclass[preprint,aps,amssymb,showpacs,superscriptaddress,nofootinbib]{revtex4}
\usepackage{accents}
\usepackage{bbold}
\usepackage{graphicx}
\newcommand{\Case}[2]{{\textstyle \frac{#1}{#2}}}

\begin{document}
\preprint{IMSc/2012/11/17}

\title{Polymer quantization and Symmetries}

\author{Ghanashyam Date}
\email{shyam@imsc.res.in}
\affiliation{The Institute of Mathematical Sciences\\
CIT Campus, Chennai-600 113, INDIA.}
\author{Nirmalya Kajuri}
\email{nirmalya@imsc.res.in}
\affiliation{The Institute of Mathematical Sciences\\
CIT Campus, Chennai-600 113, INDIA.}

\begin{abstract} 
Polymer quantization was re-discovered during the construction of Loop
Quantum Cosmology. For the simplest quantum theory of one degree of
freedom, the implications for dynamics were studied for the harmonic
oscillator as well as some other potentials. For more degrees of
freedom, the possibility of continuous, kinematic symmetries arises.
While these are realised on the Hilbert space of polymer quantum
mechanics, their infinitesimal versions are not supported. For an
invariant Hamiltonian, these symmetry realizations imply infinite
degeneracy suggesting that the symmetry should be spontaneously or
explicitly broken. The estimation of symmetry violations in some cases
have been analysed before. Here we explore the alternative of shifting
the arena to the distributional states. We discuss both the polymer
quantum mechanics case as well as polymer quantized scalar field.
\end{abstract}

\pacs{04.60.Pp, 04.60.Kz, 98.80.Jk}

\maketitle

\section{Introduction}
In Loop Quantum Gravity \cite{LQGRefs}, the twin demands of SU(2) gauge
invariance and diffeomorphism covariance, are met by the use of the
holonomies as basic variables and an inner product defined using the
Haar measure on SU(2).  The Hilbert space one gets is unique up to
unitary equivalence \cite{Uniqueness}. This also has the unusual feature
that while the diffeomorphisms have well defined unitary action, their
infinitesimal versions cannot be defined as operators \cite{JMP95}. When
specialized to the mini-superspace models, analogous procedure leads to
the so-called {\em polymer quantization}\footnote{We thank an anonymous
referee for pointing out the first two references in
\cite{Shadow}.}\cite{Shadow}. Its Hilbert space is {\em non-separable}
and has the same feature of finite translations being well defined but
not the infinitesimal generators - the momenta variables. The Stone-von
Neumann Theorem is evaded by relaxing the requirement of (weak)
continuity of the representation of the Weyl-Heisenberg algebra. One can
however introduce an {\em approximate} version of momenta and construct
corresponding non-relativistic dynamics. This necessarily introduces a
fundamental scale and modifies the energy spectra. Nevertheless for
certain systems, it can be seen explicitly that the deviations from the
usual Schrodinger quantized model are essentially indistinguishable
observationally. Such generic conclusions have been obtained for the one
dimensional harmonic oscillator \cite{Shadow} and inverse power
potentials in (effectively) one dimension \cite{InversePotl}. For
particles moving in more dimensions, we have the possibilities of
rotationally invariant systems and a natural question is to ask how the
symmetry can be incorporated.  This question, in the more general
context of Galilean symmetries has been addressed by Dah-Wei Chiou
\cite{DahWeiChiou}. He also noted that while finite group actions are
well defined, the infinitesimal ones are not. He then explored the
`approximated forms' of the usual generators (which  do not form a
closed algebra) and concluded that the deviations are small within the
domain of validity of the non-relativistic model.

We would like to ask if this technical feature of non-existence of
infinitesimal generators has any physically relevant consequences? Is
this necessarily an undesirable feature? If yes, how is the role of the
polymer representation to be understood? After all in the LQG context,
the analogue of polymer representation is very much physically well
motivated but forms only an intermediate step due to the constrained
nature of the system.

The short answers are that the non-existence of infinitesimal generators
of rotations in the polymer representation {\em also implies infinite
degeneracy for any rotationally invariant Hamiltonian} and this is a
physically undesirable feature. One option then is to break the symmetry
either explicitly or spontaneously.  Yet another possibility is to note
that we could view the polymer Hilbert space as part of a Gelfand-like
triple, Cyl $\subset H_{\mathrm{poly}} \subset$ Cyl$^*$ as in the case
of LQG and define infinitesimal generators on a suitable {\em subspace}
of Cyl$^*$.  The usual Schrodinger quantization can then be recovered,
albeit trivially.  The corresponding steps in the context of a polymer
quantized scalar field reveal further possibilities. {It is the
second alternative that is explored in this work.}

In section \ref{SphericalSymmetry} we briefly present the basic
definitions of polymer quantization in terms of a triple, as well as
specify the action of rotations. We point out how rotationally invariant
Hamiltonians can be constructed and show that the spectra of such
Hamiltonians are infinitely degenerate. 

In section \ref{DualActions} we show how to define infinitesimal
generators on a suitable subspace of the dual Cyl$^*$. A new inner
product can be naturally defined on this subspace which makes these
generators self-adjoint and also makes the completion unitarily
equivalent to the Schrodinger quantization. Although recovering
Schrodinger quantization is hardly the aim, we view this as an
illustration of a {\em multi-step} quantization procedure which could be
needed in more complex systems. 

With this in view, polymer quantized scalar field is considered in the
section \ref{ScalarField}. Although it shares the features seen in
polymer quantum mechanics, there seem to be many more possibilities for
a quantum theory admitting infinitesimal symmetries.  In the last
section \ref{Discussion} we give a summary and conclude with a
discussion. 

{We would like to emphasize the viewpoint taken in this work.  While the
polymer quantization, especially in the field theory context, is
naturally adapted to diffeomorphism covariance, nothing prevents us from
using it in the context of a fixed background geometry and coordinates.
The background structures limit the diffeomorphisms to isometries of the
background geometry and now become {\em symmetries} (transformations
among physical states leaving the dynamics invariant). We are concerned
with the representations of these symmetries in the Hilbert space of
polymer quantization. The potential violations, if any, refer to these
symmetries and {\em never} to any local invariances (gauge invariances).
With this understood, we work with a background geometry which is
Euclidean and explore the implications of polymer quantization with
regards to symmetry implementation.  }

\section{Rotational Invariance in Polymer Quantum Mechanics}
\label{SphericalSymmetry}

Consider a non-relativistic particle moving in three dimensions.
Classically it is described by the configuration space, $\mathbb{R}^3$
coordinatized by $\vec{q} \leftrightarrow q^i, i = 1, 2, 3.$ To
construct Polymer quantization, choose a countable set, $\gamma$, of
3-dimensional vectors $\vec{k}_j$ and define a set Cyl$_{\gamma}$ of
linear combinations of functions of $\vec{q}$ of the form: Cyl$_{\gamma}
:= \{\sum_{j} f_j e^{i \vec{k}_j\cdot\vec{q}}, f_j \in \mathbb{C} \}$.
Here the coefficients $f_j$ satisfy certain regularity conditions
\cite{Shadow} which do not concern us here.  Next, define the set of
functions of $\vec{q}$, Cyl := $\cup_{\gamma}$Cyl$_{\gamma}$.  On this
set, define the inner product\footnote{Strictly, it is not necessary to
give an explicit expression for the inner product. In fact, analogous
expression cannot be given when one wants to realise Lorentz symmetry.
It is enough to stipulate the orthonormal set, as in
eq.(\ref{Orthonormality}).},
\begin{eqnarray}
\langle \psi|\psi'\rangle & := & \lim_{R \to \infty}\frac{3}{4\pi
R^3}\int_{0}^{R}q^2dq\sin{\theta}d\theta d\phi \
\psi^*(\vec{q})\psi'(\vec{q}) ~~~ \Rightarrow \\
\langle \vec{k}|\vec{k}'\rangle & := & \lim_{R \to \infty}\frac{3}{4\pi
R^3}\int_{0}^{R}q^2dq\sin{\theta}d\theta d\phi \ e^{i(\vec{k}' -
\vec{k})\cdot\vec{q}} ~~~ = ~~~ \delta_{\vec{k}, \vec{k}'} ~~~ \forall
~~ \vec{k}, ~ \vec{k}' \in \mathbb{R}^3 \ . \label{Orthonormality}
\end{eqnarray}
Clearly, $\{e^{i\vec{k}\cdot\vec{q}}~/ \vec{k} \in \mathbb{R}^3\}$  form
an uncountable, orthonormal set and we denote them as the kets
$|\vec{k}\rangle$. We denote the completion of Cyl w.r.t. this inner
product, as $H_{\mathrm{poly}} := \overline{\mathrm{Cyl}}$.  The Hilbert
space is non-separable and we also have the natural triple, Cyl $\subset
$H$_{\mathrm{poly}} \subset$ Cyl$^*$, where Cyl$^*$ denote the algebraic
dual of Cyl. Notice that, the integration measure is invariant under
three dimensional rotations and preserves the orthonormality in the
second equation above. This will permit unitary representation of
rotation group (eqn. \ref{RotnAction} below). 

It is clear that $\vec{q}$ {\em cannot} be represented on the polymer
Hilbert space as a multiplicative operator since $q^i$ {acting on} a
basis element does not produce a countable linear combination of the
{basis elements (exponentials)}. The exponentials of the form,
$e^{ilq^j}$ however do form multiplicative (and {\em unitary})
operators. The derivatives too act invariantly on Cyl and $p_i := -
i\hbar\Case{\partial}{\partial q^i}$ are self-adjoint operators
representing the momenta. The exponentials are the eigenfunctions of the
momenta: $\hat{p}_i|\vec{k}\rangle = \hbar k_i|\vec{k}\rangle$. 

That the self-adjoint position operators $\hat{q}^i$ do not exist can be
seen more formally as well. Consider a 1-parameter family of unitary
operators, defined by ${\cal U}(\alpha, \vec{m})|\hat{k}\rangle :=
|\hat{k} + \alpha \vec{m}\rangle ~ \forall ~ \vec{k} \in \mathbb{R}^3$.
For any vector $\vec{\ell}$, $\langle\vec{\ell}|{\cal U}(\alpha,
\vec{m})|\vec{\ell}\rangle = \langle\vec{\ell}|\vec{\ell} + \alpha
\vec{m}\rangle = \delta_{\alpha, 0}$, as implied by the orthonormality.
Hence, the family of unitary operators is {\em not} weakly continuous at
$\alpha = 0$.  {\em If} a self-adjoint operator of the form
$\vec{m}\cdot\vec{q}$ existed, then we could define a one parameter
family of unitary operators ${\cal V}(\alpha, \vec{m}) :=
e^{i\alpha\vec{m}\cdot\vec{q}}$ which {\em is continuous at $\alpha =
0$} and precisely matches the ${\cal U}(\alpha, \vec{m})$ family, thus
reaching a contradiction. Hence, on the polymer Hilbert space, the
momenta and exponentials of positions are well defined operators but
there are {\em no} self-adjoint operators representing positions. This
feature of polymer quantization has profound implications for
implementation of continuous, non-abelian symmetries.

Recall, that any group of symmetries is represented in a quantum theory
by {\em unitary} operators\footnote{We will not be considering time
reversal or charge conjugation symmetries, so we will not consider
anti-unitary operators.}, with the states transforming as $|\psi\rangle
\to |\psi_g\rangle := U(g)|\psi\rangle$ and the operators transforming
as, $A \to A_g := U(g)A\, U^{\dagger}(g)$ for each group element $g \in
{\cal G}$. The specific unitary operators representing specific symmetry
operation can be determined by {\em stipulating how the basic
observables transform}. For example, with $q^i, p_i$ being the basic
observables in the usual quantization, the unitary operators
corresponding to rotations, are determined by:
\begin{equation}
q^i_{\Lambda} := U(\Lambda)q^i\, U(\Lambda)^{\dagger} =
\Lambda^i_{~j}q^j ~~~,~~~ p_i^{\Lambda} := U(\Lambda)p_i\,
U(\Lambda)^{\dagger} = \Lambda^j_{~i}p_j ~~~,~~~
\Lambda^i_{~m}\Lambda^j_{~n}\delta^{mn} = \delta^{ij}
\end{equation}

For infinitesimal rotations, $\Lambda^i_{~j} := \delta^i_{~j} +
\epsilon^i_{~j}\, , ~U(\mathbb{1} + \epsilon) := \mathbb{1} -
\Case{i}{\hbar} \epsilon\cdot \hat{J}$ we get,
\begin{equation}
-~\frac{i}{\hbar}[\epsilon\cdot \hat{J}, q^i] ~= ~ \epsilon^i_{~j}q^j
~~~,~~~ -~\frac{i}{\hbar}[\epsilon\cdot \hat{J}, p_i] ~= ~
\epsilon^j_{~i}p_j .
\end{equation}
With the identifications $\epsilon^i_{~j} := \epsilon_k {\cal
E}^{ki}_{~~j}\, , ~ \epsilon\cdot\hat{J} := \epsilon_k\hat{J}^k$, we
deduce $\hat{J}^k := {\cal E}_m^{~nk}q^m p_n$ as the operators
representing the infinitesimal generators. 

Alternatively, the operators $U(\Lambda)$ could also be determined by
specifying their action on {\em wavefunctions} - explicit functions on
the {\em configuration space} (say), eg. $\Psi_{\Lambda}(\vec{q}) :=
\Psi(\overrightarrow{\Lambda q})$.

For the polymer quantization, the defining stipulations for the action
of rotations are:
\begin{equation} \label{RotnAction}
\left(e^{i\vec{k}\cdot\vec{q}}\right)_{\Lambda} :=
U(\Lambda)\left(e^{i\vec{k}\cdot\vec{q}}\right)\, U(\Lambda)^{\dagger} =
\left(e^{ik_i\Lambda^i_{~j}q^j}\right) ~~~,~~~ p_i^{\Lambda} :=
U(\Lambda)p_i\, U(\Lambda)^{\dagger} = \Lambda^j_{~i}p_j 
\end{equation}

Noting that $|\hat{k}\rangle$ are eigenstates of $\hat{p}_i$, it
follows,
\begin{eqnarray}
U^{\dagger}(\Lambda)\hat{p}_i U(\Lambda)|\vec{k}\rangle & = &
(\Lambda^{-1})^j_{~i}\hat{p}_j |\vec{k}\rangle  ~ = ~
(\Lambda^{-1})^j_{~i}k_j|\vec{k}\rangle \nonumber \\
\therefore \hat{p}_i \left[U(\Lambda)|\vec{k}\rangle\right] & = &
\left[(\Lambda^{-1})^j_{~i}k_j\right]\left[U(\Lambda)|\vec{k}\rangle\right]
\nonumber \\
\therefore U(\Lambda)|\vec{k}\rangle & = &
|(\Lambda^{-1})^j_{~i}k_j\rangle
\end{eqnarray}
Evidently, this action of rotation group on the polymer Hilbert space is
{\em reducible}, with the orbit through any $\vec{k}$ being spanned by
the orthonormal kets $\{|\vec{k}'\rangle\}$ with $\vec{k}'$ lying on the
2-sphere through $\vec{k}$. The {\em subspace} spanned by $\{
|\vec{k}\rangle , \vec{k}\cdot\vec{k} = \mbox{constant}\}$, forms an
{\em irreducible representation} and is clearly infinite dimensional.

{This may come as a surprise as one recalls the theorem that all
unitary, irreducible representations of the rotation group (indeed any
compact group) are finite dimensional.  However it is to be noted that
the theorem is proved only for {\em continuous} representations of the
group (which arise from and also induce, representations of the
corresponding Lie algebra). It is also a theorem that {\em if $G$ is a
locally compact topological group whose every irreducible representation
on a Hilbert space is continuous, then the group itself is
discrete}\cite{BarutRaczka}. Since the rotation group is locally compact
and is not a discrete group, it must have discontinuous representations
as well and these do not have to be finite dimensional. What we have is
an explicit example of just such a representation whose discontinuous
nature is shown below.}

The above action of the rotation group, coupled with the fact that the
kets $|\vec{k}\rangle$ are ortho{\em normalised}, implies that
$U(\Lambda)$ also {\em cannot} be weakly continuous. Unlike the one
dimensional case where the group action necessarily transformed a basis
vector to another basis vector, here we have the possibility that
$\vec{k}$ could be along the axis of rotation represented by
$U(\Lambda)$ and hence invariant under $U(\Lambda)$. To show
discontinuity, consider any one parameter subgroup of rotations. All
these rotations will leave some particular axis invariant.  Choose any
$\vec{k}$, orthogonal to this axis. Now the subgroup action transforms a
basis ket to another distinct basis ket.  The lack of weak continuity
for every 1-parameter subgroup follows as before and we cannot write
$U(\Lambda) = \mathbb{1} - \Case{i}{\hbar}\epsilon\cdot\hat{J}$. Note
that it is {\em not} the case that {\em every} one parameter family of
unitary operators is necessarily non-continuous. For continuity to be
possible, members of the unitary family must {\em not} map any {\em
basis vector to another basis vector}.

So, while we do not have representation of infinitesimal action, finite
rotations are perfectly well defined. However for rotations to be a {\em
symmetry}, their action must also preserve the dynamics. Classically, we
have three `elementary' rotational invariants: $p\cdot p$, \,$q\cdot q$
and $p\cdot q$. Only first of these can be promoted to operator on the
polymer Hilbert space. A Hamiltonian which is a function of $p^2$ alone
will describe only a `free' dynamics.  Is this the only possible
rotationally invariant dynamics supported by the polymer Hilbert space?
Not quite. As noticed in the context of the `improved quantization' of
LQC, exponentials of arbitrary functions of momenta, times $q^i$ (i.e.
functions {\em linear} in $q^i$) can also be promoted to well defined
operators\footnote{We thank Alok Laddha for pointing this out.}.  This
is because the Hamiltonian vector field $X_{q^i}$ generates translations
along $p_i$ and any function of $\vec{p}$ multiplying $X_{q^i}$
generates more general infinitesimal transformations, also along $p_i$.
While $X_{q^i}$ cannot be promoted to an operator, its exponential which
generates {\em finite diffeomorphisms} can be!  Incorporating rotational
invariance, we can thus have unitary operators of the form $e^{\pm
if(p^2)\,p_iq^i}$.  From these, the corresponding $sin$ and $cos$
self-adjoint operators can be defined.  A candidate rotationally
invariant Hamiltonian will be a function of $p^2$ and the $sin, cos$
operators. There is no corresponding trick to use the $q\cdot q$
invariant.

To compute the action of finite diffeomorphism, say by unit parameter,
consider the integral curves defined by,
\begin{eqnarray}\label{Scale}
\frac{d p_i}{d\lambda} ~ = ~ f(p\cdot p)\,p_i & \Rightarrow & \frac{d
p\cdot p}{d\lambda} = 2 (p\cdot p)f(p\cdot p) \nonumber \\
\int_0^1 d\lambda & = &
\frac{1}{2}\int_{p^2_{\mathrm{initial}}}^{p^2_{\mathrm{final}}}\frac{dp^2}{p^2
f(p^2)}
\end{eqnarray}
This defines the change in the $p\cdot p$ for unit change in the
parameter. Notice that the vector field is {\em radial}, and therefore
the integral curves are in the radial direction (in `p'-space) and for
unit change in the parameter, connect two spheres of radii
$p^2_{\mathrm{initial}}$ and $p^2_{\mathrm{final}} := \xi^2
p^2_{\mathrm{initial}}$. The corresponding unitary operator is then
defined by,
\begin{equation}
\widehat{e^{-if(p^2)p\cdot q}}|\vec{k}\rangle := |\vec{k}' =
\xi\vec{k}\rangle ~,
\end{equation}
the scale $\xi$ being determined by eq.(\ref{Scale}).

Thus, we {\em can} have non-trivial rotationally invariant dynamics.
However, there is now a different problem. As noted before, the unitary
representation of SO(3) on the polymer Hilbert space is reducible with
irreducible representations carried by ${\cal H}_{\sigma} :=
\mathrm{span}\{|\vec{k}\rangle\, , k\cdot k = \sigma^2 > 0\}$. Each of
these is infinite dimensional. Each eigenspace of any invariant
Hamiltonian will carry a representation of SO(3) which has to be
infinite dimensional, being made up of some of the irreducible
representations together possibly with the trivial representation
($\sigma = 0$).  Thus we face the problem of {\em infinite degeneracy}
which is physically untenable: the partition function of such a system
will be undefined. We have now two possibilities: (a) rotations cease to
be a symmetry (explicit breaking of symmetry) or (b) {\em spontaneous
breaking} of rotational symmetry. 

To see both possibilities, we first seek an approximate substitute for
the position operators.  The operators, $e^{i\vec{k}\cdot\vec{q}}$,
allow us to define families of self-adjoint operators. For instance,
choosing $\vec{k}_j := \delta\hat{e}_j, ~ \hat{e}_j$ a unit vector, we
can define $\mathrm{sin_{\delta \hat{e}_j} :=
(2i)^{-1}(e^{i\delta\hat{e}_j\cdot\vec{q}} -  e^{-
i\delta\hat{e}_j\cdot\vec{q}})}$ and a $\mathrm{cos}$ operator
analogously\footnote{These operators however do not suffice to represent
the Lie algebra of rotations \cite{DahWeiChiou}.}. We could choose
several triplets of linearly independent unit vectors $\hat{e}_j$ and
also choose many different parameters $\delta$'s (equivalently, finitely
many $\vec{k}_j$).  If we collect finitely many of such sets and
restrict ourselves to observables which are functions of these (and the
momentum) operators, then from any given $|\vec{k}_0\rangle$, we will
generate a collection of basis vectors, $\{|\vec{k}_0 + \sum_j
n_j\vec{k}_j\rangle, n_j \in \mathbb{Z}\}$.  The closed subspace
generated by this set will be a {\em proper subspace} of the polymer
Hilbert space and is clearly {\em separable}. If we also include
operators which are exponentials in $p\cdot q$, discussed above, then
the lattice generated will also involve scaling determined by the
choices for $f(p^2)$. As long as the number of such operators is finite,
we will continue to have separable sectors. The chosen set of
observables, will act invariantly on each of these subspaces and will
provide {\em superselection sectors}. Observe that among the chosen
class of observables, we can also have an invariant Hamiltonian. Action
of rotations however mixes different sectors and we have {\em
spontaneous breaking of rotational invariance}.  If we chose a
Hamiltonian involving the approximated position operators, we have {\em
explicit breaking} of rotations controlled by the $\delta$-parameter(s).
The example of spherically symmetric harmonic oscillator in three
dimensions illustrates this. For a economical parametrization of
violation, we can choose a single common $\delta$. For sufficiently
small values of this, at a certain level of observational precision, it
is of course possible to have the illusion of rotational invariance.

To summarise, having made a choice of the polymer Hilbert space
$H_{\mathrm{poly}}$, we {\em can} have exact rotational symmetry with a
some what restricted form of dynamics (no $q\cdot q$ dependence) but
with uncountably infinite degeneracy. To avoid the problem of infinite
degeneracy, the symmetry must be broken - either explicitly {\em or}
spontaneously. By introducing separable sectors we can see both
possibilities.

One can however also view the polymer quantization as an intermediate
step in constructing a quantum theory, much as the kinematical Hilbert
space of LQG is. Using the triple, we can try to select a suitable {\em
subspace} of Cyl$^*$ on which infinitesimal generators can be defined.
With a suitable choice of a new inner product, we can obtain a
`physical' Hilbert space with a rotationally invariant dynamics.
\section{Infinitesimal Generators} \label{DualActions}
The possibility of looking to Cyl$^*$ for a home to a suitable quantum
theory is inspired by analogous steps taken in the context of LQG. In
LQG, the step is motivated for a very different reason. The kinematical
quantization is essentially forced upon us by the demand of SU(2)
invariance and diffeomorphism covariance. Since there are constraints
whose kernels are in general distributional, an appropriate
diffeo-invariant subspace of corresponding Cyl$^*$ is a natural arena.
In our case, the polymer quantization itself is not a compulsion, but is
a useful illustration of a {\em multi-step construction of a quantum
theory.}

Recall that construction of $\mathrm{H_{poly}}$ naturally gave us the
triple $\mathrm{Cyl \subset H_{poly} \subset Cyl^*}$. This structure
provides us with a convenient representation of the elements $(\Psi|$ of
$\mathrm{Cyl^*}$ by complex valued linear functions
$\mathrm{\psi^*(\vec{k}) := (\Psi|\vec{k}\rangle}$. No smoothness
properties are assumed at this stage for these functions. Furthermore,
for every operator A : Cyl $\to$ Cyl, we can define an operator
$\mathrm{\tilde{A}: Cyl^* \to Cyl^*}$ by the `dual action', eg.
$\mathrm{(\tilde{A}\Psi|f\rangle := (\Psi|Af\rangle, \forall~ |f\rangle
\in Cyl, ~ \forall~ (\Psi| \in Cyl^*}$. Conversely, given an operator
$\mathrm{\tilde{A}}$ defined on {\em all} of Cyl$^*$, we can define an
operator A on Cyl by the same equation as above (read backwards).  In
particular this means that we have the operators
$\mathrm{\tilde{U}(\Lambda)}$ defined on Cyl$^*$. We will use these to
define infinitesimal generators on Cyl$^*$. We will also define the
position operators.

We begin with infinitesimal rotation generators.
\begin{eqnarray}
\mathrm{(\Psi|U(\mathbb{1} + \epsilon) - U(\mathbb{1} -
\epsilon)|\vec{k}\rangle} 
& = & \mathrm{ (\Psi|\vec{k}} +
\overrightarrow{\epsilon\mathrm{k}}\rangle - (\Psi|\vec{k}
- \overrightarrow{\epsilon\mathrm{k}}\rangle \nonumber \\
& \approx & \mathrm{2 \epsilon_l{\cal E}^{li}_{~~j}k^j\frac{\partial
\psi^*}{\partial k^i}} \nonumber \\
\therefore \mathrm{\lim_{\epsilon_l \to 0}(\Psi|\frac{U(\mathbb{1} +
\epsilon)
- U(\mathbb{1} - \epsilon)}{2\epsilon_l}|\vec{k}\rangle}
& = & \mathrm{{\cal E}^{li}_{~~j}k^j\frac{\partial \psi^*}{\partial
k^i}} \nonumber \\
\therefore ( \mathrm{J}^\mathrm{l} \Psi|\vec{\mathrm{k}}\rangle & := &
\mathrm{-i\hbar}{\cal E}^{\mathrm{li}}_{~~\mathrm{j}}\
\mathrm{k}^{\mathrm{j}}\ \frac{\partial\psi^*}{\partial
\mathrm{k}^\mathrm{i}}
\end{eqnarray}

Notice that these operators are defined only on a {\em subspace} of
Cyl$^*$, consisting of those $(\Psi|$ whose corresponding
$\psi^*(\vec{\mathrm{k}})$ are {\em differentiable functions}.  Hence,
by dual action we {\em cannot} define the corresponding operators on
Cyl. 

Next, recall the sin$_{\delta\mathrm{\hat{e}_j}}$ operators defined in
the previous section.  For each orthonormal triad, $\hat{e}_j, \ j = 1,
2, 3,~ \hat{e}_i\cdot\hat{e}_j = \delta_{ij}$ and a small parameter
$\delta$, we have, $U_{\delta \hat{e}_j}(\vec{q}) :=
e^{i\delta\hat{e}_j\cdot\vec{q}}$ and $ \mathrm{sin}_{\delta\hat{e}_j} ~
:= ~ (2i)^{-1}({U_{\delta \hat{e}_j}(\vec{q}) - U_{-\delta
\hat{e}_j}(\vec{q})})~.  $ Now,
\begin{eqnarray}
\mathrm{2i (\Psi|sin_{\delta\hat{e}_j}|\vec{k}\rangle} 
& = & \mathrm{ (\Psi|\vec{k} + \delta\hat{e}_j\rangle - (\Psi|\vec{k}
- \delta\hat{e}_j\rangle} \nonumber \\
& = & \mathrm{\psi^*(\vec{k} + \delta\hat{e}_j) - \psi^*(\vec{k} -
\delta\hat{e}_j) }\nonumber \\
& \approx & \mathrm{2 \delta \hat{e}_j \frac{\partial \psi^*}{\partial
\vec{k}}} \nonumber \\
\therefore \mathrm{\lim_{\delta \to 0}(\Psi|\frac{sin_{\delta
\hat{e}_j}}{\delta}|\vec{k}\rangle} & = & \mathrm{- i
\hat{e}_j\cdot\vec{\nabla}_{\vec{k}}\psi^* }
\end{eqnarray}
Thus, by restricting to functions $\psi^*$ which are at least
differentiable, we can define a {\em position operator} on a {\em
subspace} of Cyl$^*$ via the dual action, 
\begin{equation}
\mathrm{(\hat{e}_j\cdot\vec{q}\ \Psi} |\vec{\mathrm{k}}\rangle ~ := ~
\mathrm{- i \hat{e}_j\cdot\vec{\nabla}_{\vec{k}}\psi^* ~~ , ~~ \forall ~
(\Psi| \in Cyl^* ~~ \mbox{such that} ~ \psi^*(\vec{k}) \mbox{ is
differentiable}}\ .
\end{equation}

It is easy to see that the position operators defined above and the
momentum operators defined by dual action, also satisfy,
\[
\mathrm{(~ [\hat{e}_m\cdot\vec{q}\ , \ \hat{e}_n\cdot\vec{p}]~
\Psi|\vec{k}\rangle ~ = ~
(~\{i\hbar\hat{e}_m\cdot\hat{e}_n\}~\Psi|\vec{k}\rangle} \ .
\]

{So far we have not specified any subspace of Cyl$^*$ except to say that
it should consists of, at least, differentiable functions. The space of
all differentiable functions is too large a subspace to choose.  We are
guided in our choice of a subspace by the requirement that the
`position' and the `momentum' operators be self-adjoint with respect to
a suitable inner-product and satisfy the canonical commutation relation
on an invariant, common dense domain. Representations of the canonical
commutation relations are usually analyzed by going to the bounded,
unitary operators (exponentials of the positions and momenta) ,
satisfying the Weyl-Heisenberg relations. The Stone-von Neumann theorem
then guarantees a unique continuous representation of the
Weyl-Heisenberg relations and the corresponding canonical commutation
relations. This representation corresponds to the choice of Schwartz
space as the subspace of Cyl$^*$ and the usual inner product with the
Lebesgue measure. The Hilbert space is then obtained by completing
Schwartz space in the $L_2$ norm. Making this choice, we just get back
the usual Schrodinger quantization using functions of ``momenta'',
$\vec{k}$ instead of functions of ``positions'', $\vec{q}$.  The
intermediate polymer quantization has only led us to the Heisenberg
representation instead of the Schrodinger representation. The measure
being invariant under rotations also admits (unitary) representation of
infinitesimal rotations. It is interesting to note that one can also
choose a subspace which is {\em larger}, eg space of $\psi^*(\vec{k})$
which are normalizable with respect to a Sobolev norm, and choose the
Lorentz invariant measure $\Case{d^3k}{2\sqrt{\vec{k}\cdot\vec{k} +
m^2}}$, to construct Hilbert space of a {\em free, relativistic particle
of mass $m$}\cite{Jaffe}. This is not our primary concern though.}

This is obviously a roundabout way of arriving at the usual
quantization. But it shows that (a) not every choice of quantization may
be flexible enough for physical modeling and (b) we can reach a
satisfactory quantum theory by modifying the quantization algorithm. In
principle, if the quantum theory constructed from a subspace of Cyl$^*$
were {\em not} satisfactory, we could repeat the process forming a new
triple. This is further discussed in the last section. In the next
section we discuss the case of a scalar field theory.
\section{The case of a Scalar Field theory}\label{ScalarField}
Can rotational invariance be supported in a `polymerised scalar field
theory'?  Consider the example of a scalar field $\phi(\vec{x})$ defined
on $\mathbb{R}^3$. The rotations act on the space which in turn induces
a transformation on the field: $\phi'(\vec{x}) =
\phi(\overrightarrow{\Lambda x})$. The polymer quantization of the
scalar field is done as follows \cite{PolymerScalar}.

Define a vertex set $V = (\vec{x}_1, \vec{x}_2, \ldots, \vec{x}_n)$, of
finitely many, {\em distinct} points. For {\em non-zero} real numbers
$\lambda_j, j = 1, \ldots, n$, define the functions ${\cal
N}_{V,\vec{\lambda}}(\phi) := e^{i\sum_j \lambda_j \phi(\vec{x}_j)}$.
For each fixed set $V$, let Cyl$_V$ denote the set of finite, complex
linear combinations of these functions. Let Cyl := $\cup_V$ Cyl$_V$.
Thus every element of Cyl is a function of $\phi$ which is a {\em
finite} linear combination of functions ${\cal N}_{V, \vec{\lambda}}$
for {\em some} vertex set $V$ and {\em some} choice of $\vec{\lambda}$.
Define an inner product, 
\begin{eqnarray}
\langle \psi | \psi'\rangle & := & \int d\mu(\phi)
\psi^*(\phi)\psi'(\phi) = \int d\mu(\phi) \sum_{V,\vec{\lambda}, V',
\vec{\lambda}'} C^*_{V,\vec{\lambda}}C'_{V',\vec{\lambda}'}
e^{i\sum_k\lambda'_k\phi(\vec{x}'_k) - i\sum_j\lambda_j\phi(\vec{x}_j)}
\label{CylInnerProduct} 
\end{eqnarray}

Observe that each term in the summand is again of the form ${\cal
N}_{V\cup V', \vec{\lambda}, \vec{\lambda}'}$, except that all vertices
in the union $V\cup V'$ are not necessarily distinct. If $\vec{x}'_k =
\vec{x}_j$, then the exponent would be $(\lambda'_k -
\lambda_j)\phi(\vec{x}_j)$. If the $\lambda's$ are equal, then the
exponent is identically zero and the integral contributes to the sum.
Otherwise, the integral gives zero. It follows that ${\cal
N}_{V,\vec{\lambda}}$ and ${\cal N}_{V', \vec{\lambda}'}$ are orthogonal
unless the two sets of vertices coincide {\em and} their corresponding
$\lambda$'s are equal. 

The Hilbert space H$_{\mathrm{poly}}$, is obtained as the Cauchy
completion of Cyl with respect to  this inner product. The functions
${\cal N}_{V,\vec{\lambda}}(\phi)$, with {\em every} $\lambda \neq 0$,
form an orthonormal basis for the polymer Hilbert space. The constant
function corresponding to empty vertex set, ${\cal N}(\phi) = 1$, is
also included in the basis.

Action of rotations on Cyl is defined by $[U_{\Lambda}\psi](\phi) :=
\psi(\Lambda\circ \phi)$. Evaluating it on the elementary functions lead
to,
\begin{equation} \label{RotActionOnBasis}
{\cal N}_{V,\vec{\lambda}}(\phi) ~ \to ~ {\cal N}'_{V',
\vec{\lambda}'}(\phi) ~ :=~  {\cal N}_{V, \vec{\lambda}}(\phi') ~ =~
{\cal N}_{V', \vec{\lambda}}(\phi)
\end{equation}
The middle equality is the definition of the action, $\phi' =
\Lambda\circ\phi$ and we have used the scalar nature of $\phi$,
$\phi'(\vec{x}) = \phi(\overrightarrow{\Lambda x})$, in the last
equality.

Observe that under the action of rotation $\Lambda$, a vertex set $V =
(\vec{x}_1, \ldots, \vec{x}_n)$ changes to a new vertex set $V' :=
(\overrightarrow{\Lambda x_1}, \ldots, \overrightarrow{\Lambda x_n})$.
The $\lambda's$ are unchanged and the field is evaluated at the
transformed points.  Since the $\lambda$'s do not change and the inner
product depends only on them, the inner product among elementary
functions is invariant under the action of the rotations and therefore
rotations are represented unitarily on the Hilbert space.

That this unitary action is also non-weakly-continuous can be seen
easily. For non-trivial rotation, a diagonal matrix element between
basis states is zero while for the identity rotation, the matrix element
is 1. Thus, infinitesimal generators have no representation on the
polymer Hilbert space.

The momenta variables are defined as,
\begin{equation}
P_g := \int d^3x g(\vec{x})\pi_{\phi}(x) ~ = ~ -i \hbar \int d^3x
g(\vec{x})\frac{\delta~~~~~}{\delta\phi(\vec{x})}
\end{equation}
Here $g(\vec{x})$ is a `suitably smooth' function ($\pi_{\phi}$ has
density weight 1, though not relevant here). It is easy to see that,
\begin{equation}\label{PDefn}
P_g{\cal N}_{V, \vec{\lambda}} ~ = ~ \left[\hbar \sum_j \lambda_j
g(\vec{x}_j)\right] {\cal N}_{V, \vec{\lambda}} ~~,~~ \left[ P_f, P_g
\right] = 0 ~~,~~ P_f^{\dagger} ~=~ P_f ~ .
\end{equation}
Thus the momentum representation exists and the elementary functions
${\cal N}_{V, \vec{\lambda}}$ are simultaneous eigenstates of the
momenta variables $P_g$. Under the action of rotation, $U(\Lambda)$, the
momentum variables transform as,
\begin{eqnarray}
U_{\Lambda}P_g(\pi)U^{\dagger}_{\Lambda} & := & P_g(\Lambda\circ\pi)
\nonumber \\ 
& = & P_{\Lambda^{-1}\circ g}(\pi) ~~~~ (\mbox{from the definition})
~~~~\Rightarrow \\
U_{\Lambda} P_g(\pi) & = & P_{\Lambda^{-1}\circ g}(\pi) U_{\Lambda}
\nonumber
\end{eqnarray}
This is consistent with (\ref{RotActionOnBasis}).  Let us use the
notation $|V, \vec{\lambda} \rangle \leftrightarrow {\cal N}_{V,
\vec{\lambda}}(\phi)$.

Observe that $e^{i\lambda\phi(\vec{x})}$, a `{\em point holonomy
operator}', clearly acts as a multiplication operator:
\begin{eqnarray}
e^{i\lambda\phi(\vec{x})}|V, \vec{\lambda}\rangle & := & \left\{
\begin{array}{ll}
|\vec{x}_1,\ldots,\vec{x}_n,\vec{x}\
;\lambda_1,\ldots,\lambda_n,\lambda\rangle & \mathrm{if} ~ \vec{x} \neq
\vec{x}_i ~ \mathrm{for ~ any~ i}\\
|\vec{x}_1,\ldots,\vec{x}_k\ldots\vec{x}_n\ ;\lambda_1,\ldots,\lambda_k
+ \lambda,\ldots\lambda_n\rangle & \mathrm{if}~ \vec{x} = \vec{x}_k ~ ,
~ \lambda + \lambda_k \neq 0\\
|\vec{x}_1,\ldots,,\ldots\vec{x}_n\ ;\lambda_1,\ldots,,
\ldots\lambda_n\rangle & \mathrm{if}~ \vec{x} = \vec{x}_k ~ , ~ \lambda
+ \lambda_k = 0
\end{array} \right.
\end{eqnarray}
In the last equation, the $\vec{x}_k, \lambda_k$ labels are missing on
the right hand side.

What about the scalar field operator itself? It does {\em not} exist
since the point holonomy operators are not weakly continuous, exactly as
in the point particle case.  {In the usual Schrodinger type
representation too, a scalar field operator exists {\em only} as an
operator valued distribution. This has to do with the presence of Dirac
delta in the canonical commutation relations. In the polymer
representation it does not exist even as an operator valued
distribution.} 

Now consider an element $\mathrm{(\Psi| \in Cyl^*}$. Its action on an
elementary function ${\cal N}_{V, \vec{\lambda}}(\phi)$ is given by,
\[
(\Psi|V, \vec{\lambda}\rangle =: \psi^*(\vec{x}_1, \cdots, \vec{x}_n,
\lambda_1, \cdots, \lambda_n) ~~,~~ \mbox{distinct $\vec{x}\ 's$ and
non-zero $\lambda's$}.
\]
Under the action of rotations, the arguments of the elementary function
change: $|V, \vec{\lambda}\rangle \to |V', \vec{\lambda}\rangle$.
Thus, if we choose the functions $\psi^*$'s to be differentiable, we can
define infinitesimal rotations as operators on a subspace of
$\mathrm{Cyl}^*$, exactly as before. Explicitly,
\begin{eqnarray}
\mathrm{(\Psi|U(\mathbb{1} + \epsilon) - U(\mathbb{1} - \epsilon)|V,
\vec{\lambda}\rangle} 
& = & \mathrm{ (\Psi|V_+', \vec{\lambda}\rangle - (\Psi|V'_-,
\vec{\lambda} \rangle } \nonumber \\
& = & \mathrm{ \psi^*(\vec{x}_1 + \overrightarrow{\epsilon x_1}, \cdots,
\vec{x}_n + \overrightarrow{\epsilon x_n}, \vec{\lambda})} - \nonumber
\\
& & \hspace{2.0cm} \mathrm{\psi^*(\vec{x}_1 - \overrightarrow{\epsilon
x_1}, \cdots, \vec{x}_n - \overrightarrow{\epsilon x_n}, \vec{\lambda})
} \nonumber \\
& \approx & \mathrm{2 \epsilon_k{\cal E}^{ki}_{~~j}\sum_{m = 1}^n
x_m^j\frac{\partial \psi^*}{\partial x_m^i}} \nonumber \\
\therefore \mathrm{\lim_{\epsilon_k \to 0}(\Psi|\frac{U(\mathbb{1} +
\epsilon)
- U(\mathbb{1} - \epsilon)}{2\epsilon_k}|V, \vec{\lambda}\rangle}
& = & \mathrm{{\cal E}^{ki}_{~~j}\sum_{m = 1}^n x_m^j\frac{\partial
\psi^*}{\partial x_m^i}} \nonumber \\
\therefore ( J^k \Psi|V, \vec{\lambda}\rangle & := & \mathrm{-i\hbar\,
{\cal E}^{ki}_{~~j}\sum_{m = 1}^n x_m^j\frac{\partial \psi^*}{\partial
x_m^i}}
\end{eqnarray} 
Thus, by restricting to a subspace of Cyl$^*$, corresponding to suitably
differentiable functions $\psi^*(V, \vec{\lambda})$, we can define the
generator of the infinitesimal rotations.

Likewise, to define a smeared operator scalar field on $\mathrm{Cyl}^*$,
consider,
\begin{eqnarray}
(\Psi|\phi_f^\delta|V,\vec{\lambda}\rangle & := & \int d^3x
f(\vec{x})(\Psi|\frac{e^{i\delta\phi(\vec{x})} -
e^{-i\delta\phi(\vec{x})}}{2i\delta}|V, \vec{\lambda}\rangle \\
& = & \int d^3x
\frac{f(\vec{x})}{2i\delta}\left((\Psi|V,\vec{x},\vec{\lambda},
\delta\rangle - (\Psi|V,\vec{x},\vec{\lambda}, - \delta\rangle\right)
\nonumber
\end{eqnarray}
For a generic $\vec{x}$, assuming differentiability of $\psi^*$, we will
get a function of the vertices of $V$ and the corresponding $\lambda's$
together with the additional point $\vec{x}$ and the corresponding
`$\delta$' = 0. This function cannot come from any element of Cyl$^*$
acting on $|V,\vec{\lambda}\rangle$.  Hence we should avoid getting a
contribution from a generic $\vec{x}$.  If however, $\vec{x}$ coincides
with one of the vertices in $V$, then the resultant function
(derivative) {\em is} a function of $(V, \vec{\lambda})$ and we can
interpret the right hand side as a new element of Cyl$^*$ evaluated on
the basis element $|V, \vec{\lambda}\rangle$. This can be made more
precise by employing the commonly used procedure of defining the
integral by introducing a cell decomposition adapted to the `graph'
(vertices of $V$) and {\em demanding that}
\begin{equation} \label{CylStarCondn}
\left.\frac{\partial\psi^*}{\partial\lambda_j}(\vec{x}_1,\ldots,\vec{x}_n,
\lambda_1,\ldots,\lambda_j,\ldots,\lambda_n)\right|_{\lambda_j = 0} = 0
~~, \forall~ j = 1, 2, \ldots, n \ .
\end{equation}
This condition ensures that there is no contribution from cells that do
not contain a vertex of $V$ and we are led to the definition:
\begin{equation} \label{PhiDefn}
(\widetilde{\phi_f}\Psi|V,\vec{\lambda}\rangle ~:=~ \lim_{\delta \to
0}(\Psi|\phi_f^\delta|V,\vec{\lambda}\rangle ~ := ~ -i \sum_j
f(\vec{x}_j)\frac{\partial\psi^*(\vec{x}_1,\ldots,\vec{x}_n,\lambda_1,\ldots,\lambda_n)}{\partial
\lambda_j} \ .
\end{equation} 

It is now easy to verify that
\begin{eqnarray}
(\widetilde{[\phi_f, P_g]}\Psi|V, \vec{\lambda}\rangle & := &
(\widetilde{P_g}\widetilde{\phi_f}\Psi|V,\vec{\lambda}\rangle -
(\widetilde{\phi_f}\widetilde{P_g}\Psi|V,\vec{\lambda}\rangle \nonumber
\\
& = & -i\hbar\left(\sum_{j = 1}^n f(\vec{x}_j)g(\vec{x}_j)\right)\
(\Psi|V, \vec{\lambda}\rangle \ \nonumber \\
& = & (\left\{ +i\hbar\left(\sum_{j = 1}^n
f(\vec{x}_j)g(\vec{x}_j)\right)\ \right\} \Psi|V, \vec{\lambda}\rangle \
.
\end{eqnarray}
We have thus succeeded in defining the smeared versions of the field
operators $\phi_f, P_g$ in a subspace of Cyl$^*$. 

We can also verify that the infinitesimal generators $J^k$ induce
expected actions on the smeared fields operators.
\begin{eqnarray}
(\widetilde{[J^k, \phi_f]}\Psi|V,\vec{\lambda}\rangle & = &
(\widetilde{\phi_f}\widetilde{J_k}\Psi|V,\vec{\lambda}\rangle -
(\widetilde{J_k}\widetilde{\phi_f}\Psi|V,\vec{\lambda}\rangle \nonumber
\\
& = & -i\sum_{m =
1}^{N}f(\vec{x}_m)\frac{\partial\psi^*_{J_k}(V,\vec{\lambda})}{\partial\lambda_m}
- (-i\hbar){\cal E}^{ki}_{~~j}\sum_{n =
  1}^{N}x^j_n\frac{\partial\psi^*_{\phi_f}(V,\vec{\lambda})}{\partial
  x^i_n} \nonumber \\
& = & -\hbar\sum_{m,n = 1}^{N}f(\vec{x}_m){\cal
E}^{ki}_{~~j}x^j_n\frac{\partial\psi^*(V,\vec{\lambda})}{\partial x^i_n}
\nonumber \\
& & \hspace{3.0cm} + ~~ \hbar{\cal E}^{ki}_{~~j}\sum_{m,n =
1}^{N}x^j_n\frac{\partial~~}{\partial
x^i_n}\left\{f(\vec{x}_m)\frac{\partial\psi^*(V,\vec{\lambda})}{\partial\lambda_m}\right\}
\nonumber \\
& = & i\hbar\left[-i\sum_{n = 1}^N \left({\cal E}^{ki}_{~~j}
x_n^j\frac{\partial f(\vec{x}_n)}{\partial
x_n^i}\right)\cdot\frac{\partial\psi^*(V,\vec{\lambda})}{\partial\lambda_n}\right]\nonumber
\\
& = & i\hbar ~ (\widetilde{\phi_{{\cal L}_k f}}\Psi|V,
\vec{\lambda}\rangle \hspace{1.0cm},\hspace{1.0cm} {\cal L}_k f(\vec{x})
:= {\cal E}^{ki}_{~~j}x^j\frac{\partial f}{\partial x^i}
\end{eqnarray}
Similar computation can be done for commutator of $[J^k, P_g]$.

We have now identified the {\em minimal conditions}, namely
differentiability in all arguments and the condition of equation
(\ref{CylStarCondn}), on functions $\psi^*(\vec{x}_1, \ldots, \vec{x}_n,
\lambda_1, \ldots, \lambda_n)$ in order that the smeared field operators
and the infinitesimal rotation actions are well defined. Since such an
element of Cyl$^*$ can be viewed as a sequence of {\em differentiable},
complex functions defined on $(\mathbb{R}^{3n} - \mathrm{diagonal})
\times (\mathbb{R}^n - \vec{0})$ where $\mathrm{diagonal}$ is the subset
of $\mathbb{R}^{3n}$ with two or more points coinciding, we are
restricted to a subspace of Cyl$^*$. The next step is to choose a
suitable inner product on this subspace, possibly restricted further
with additional conditions. Let us denote such a subspace by Cyl$_1$.
Here we initiate first steps. For notational simplicity, let us denote
elements of Cyl$^*$ generically by underlined letters such as as
$\underline{\Psi}, \underline{\Phi}, \underline{[V, \vec{\lambda}]},
\ldots$ etc.

Heuristically, we can represent each element of Cyl$_1$ and a yet to be
defined inner product as,
\begin{eqnarray}
\underline{\Psi} & := & \sum_{V, \vec{\lambda}} \psi^*(V,
\vec{\lambda})\underline{[V, \vec{\lambda}]}\ ,
\label{FormalExpansion}\\
\langle \underline{\Psi}, \underline{\Phi} \rangle & := & \sum_{V,
\vec{\lambda}} \sum_{V', \vec{\lambda}'} \psi(V,
\vec{\lambda})\phi^*(V', \vec{\lambda}') \langle \underline{[V,
\vec{\lambda}]}, \underline{[V',\vec{\lambda}']}\rangle \nonumber \\
& := & \sum_{V, \vec{\lambda}} \sum_{V', \vec{\lambda}'} \psi(V,
\vec{\lambda})\phi^*(V', \vec{\lambda}') {\cal G}(V, \vec{\lambda}; V',
\vec{\lambda}') 
\end{eqnarray}
The coefficients $\psi^*(V,\vec{\lambda})$ in the first line, contain
the information about the subspace, Cyl$_1$. The ${\cal G}$ denotes the
inner product between `basis' elements. 

For example, Cyl is a subspace of Cyl$^*$ through the natural embedding
$|V,\vec{\lambda}\rangle \in $ Cyl $\to \underline{[V,\vec{\lambda}]}
\in $ Cyl$^*$. If Cyl$_1$ were to be this subspace, then the
$\psi^*(V,\vec{\lambda})$ in eqn.(\ref{FormalExpansion}) would be
non-zero only for finitely many $(V, \vec{\lambda})$ sets and ${\cal
G}(V, \vec{\lambda}; V', \vec{\lambda}')$ would equal $
\delta_{V,V'}\delta_{\vec{\lambda}, \vec{\lambda}'}$. The double
summation would then collapse to a {\em finite} sum over $(V,
\vec{\lambda})$ (compare eqn.(\ref{CylInnerProduct})). Likewise, if
Cyl$_1$ were to echo the Hilbert space of the r-Fock construction
\cite{RFock, PolymerScalar}, the ${\cal G}(V, \vec{\lambda}; V',
\vec{\lambda}')$ would be $ \sim \mathrm{exp}[ - \Case{1}{4}\sum_{ij}
G_{ij}(\vec{x}_i, \vec{x}_j)\lambda^i\lambda^j]$, where the sum over
$(i,j)$ is over the vertices of $V \cup V'$ and we use the notation of
\cite{PolymerScalar}. The double sum will be a finite sum since
$\psi^*(V, \vec{\lambda})$ is non-zero for finitely many $(V,
\vec{\lambda})$ sets.

More generally, we could have uncountably many non-zero $\psi^*(V,
\vec{\lambda})$ and then each $\underline{\Psi}$ can be thought of as a
potentially infinite sequence of functions, $\psi_n$, on $\sim
\mathbb{R}^{4n}$. If we choose an inner product so that the `basis
states' are orthonormal (${\cal G} \propto \delta_{V,
V'}\delta_{\vec{\lambda}, \vec{\lambda}'})$, then we may write the inner
product as,
\begin{eqnarray}
\langle \underline{\Psi}, \underline{\Phi} \rangle 
& := & \sum_{V, \vec{\lambda}} \psi(V, \vec{\lambda})\phi^*(V',
\vec{\lambda}') \\
& :\simeq & \sum_{n = 0}^{\infty}\int_{\mathbb{R}^{3n}}
d^{3n}x\int_{\mathbb{R}^{n}}d^n \lambda \ \psi(\vec{x}_1, \ldots,
\vec{x}_n, \lambda_1, \ldots, \lambda_n) \phi^*(\vec{x}_1, \ldots,
\vec{x}_n, \lambda_1, \ldots, \lambda_n) \label{HeuristicDefn} \nonumber
\end{eqnarray}
The $:\sim$ indicates that the integration measures need to be defined
and we need to put conditions to ensure convergence of the sum.

Assuming that we can choose suitable weights in the sum and measures in
the integrations, what further conditions we need to put on the
$\psi_n$'s so that our basic operators and generators are self-adjoint?
It is easy to see that we need only the usual fall-off conditions on
these so that surface terms resulting from the partial integrations drop
out. Roughly, we make each member $\psi_n$ as an element of
$L_2(\mathbb{R}^{4n})$. This indicates that it is, at least
heuristically, {\em conceivable} to choose suitable definitions to
construct a new Hilbert space. 

{Many more issues have to be addressed. Even for the point particle
case, self-adjointness and even commutation relations were not enough to
lead to a unique choice, the Weyl-Heisenberg relations are needed to be
invoked. For a field theory it is known that {\em even after invoking
the Weyl-Heisenberg relations}, there are infinitely many inequivalent
representations of the canonical commutation relations.  In the usual
case, Poincare invariance is additionally invoked to uniquely single out
the Fock representation \cite{BogoliubovEtAl}. A detailed analysis of
the possibilities is beyond the scope of the present work.} 

\section{Summary and Discussion} \label{Discussion}
We began by exploring symmetries and their violations in polymer
quantized systems. Specifically, we focused on three dimensional
rotations and explored the polymer quantized particle in three dimensions
and a scalar field defined on $\mathbb{R}^3$. It is certainly possible
to have a unitary representation of SO(3) on the polymer Hilbert space
but the representation is discontinuous and consequently does not admit
representation of its Lie algebra. The non-availability of configuration
space operators - position operators - severely restricts the possible
invariant Hamiltonians and {\em every one of these} has infinitely
degenerate eigenvalues. In effect, physically acceptable dynamics on
polymer Hilbert space {\em must necessarily violate rotational symmetry,
either explicitly or spontaneously}. In case of explicit breaking, one
can then look for economical parametrization of symmetry violations and
put bounds on the parameters. As noted in the introduction, this route
has already been followed in \cite{Shadow, DahWeiChiou, InversePotl}. We
explored another route to see if acceptable quantization, {\em with
infinitesimal symmetries}, can be arrived at viewing polymer
quantization as an intermediate step. This was done by looking for
suitable subspace(s) of the dual member of the Gelfand triple with a
hope of defining a new inner product and a new Hilbert space.  For the
point particle case we verified that it is possible to construct a new
Hilbert space which carries continuous representations of the rotation
group as well as continuous representation of the Heisenberg group. By
the Stone-von Neumann theorem, this is of course the usual Schrodinger
representation which supports the usual rotationally invariant
non-trivial Hamiltonians. The case of scalar field revealed greater
richness. There can be infinitely many choices of inner products, all of
which can support infinitesimal rotations as well as elementary smeared
field. 

In principle neither of the two routes is unnatural. It is not certain,
that continuous symmetries need be realised {\em exactly} in nature even
if observations support their existence to excellent approximation, eg
Lorentz symmetry.  Symmetries help to exercise tighter control over
theoretical frameworks but physical system may not exactly respect the
implicit idealization.  The in-built, non-invariant dynamics of a
polymer quantized system, suggests a particular parametrization of
symmetry violation eg the use of the `trigonometric' operators to build
the Hamiltonian. At least in the cases explored, such violations are
viable. 

The second alternative is anyway needed in the context of theories with
first class constraints. It could well be thought of as a multi-step
quantization procedure.  Just as in a classical theory, specified by an
action, the variables we begin with need not represent the physical
states (eg when there are constraints). However following a systematic
procedure - the Dirac algorithm of constraint analysis - we can arrive
at a formulation which is either a theory with a first class constraint
algebra or a theory without any constraints. Likewise, one could begin
with a set of basic functions on the configuration space forming a
Cyl$_0$, choose an inner product $\langle | \rangle_0$, obtain a
Cyl$^*_0$ as well as a Hilbert space H$_0$ forming a triple: Cyl$_0
\subset $H$_0 \subset$ Cyl$^*_0$.  If the model is satisfactory, we are
done. If not, look for a {\em subspace} Cyl$_1 \subset $ Cyl$^*_0$,
define a new inner product $\langle | \rangle_1$ and obtain a new triple
Cyl$_1 \subset $H$_1 \subset$ Cyl$^*_1$. Hopefully the process would
terminate after a finite number of iterations. This procedure offers a
flexibility to refine the class of observables we wish to be supported
on the quantum state space. It is constructive and could help keep the
focus on physical observables.  This possibility needs to be examined
further to see its viability/utility.

We have considered scalar field theory with `point holonomies' as basic
functions generating the commutative C$^*$ algebra.  Fermions are
similar to point holonomies as far as the label sets are concerned.  For
gauge fields, we will have the H$_{\mathrm{poly}} :=~$H$_{\mathrm{kin}}$
with the basis labeled by discrete labels. Hence the analogues of $(V,
\vec{\lambda})$ will now have embedded graphs and representation labels
of the gauge group.  One will have to impart a `manifold structure' for
these spaces of labels to attempt a definition of infinitesimal
generators in the manner discussed above. 

{
We would like to end by drawing a parallel with recent work on polymer
quantization of parametrized field theory (PFT) \cite{PFTRecent}.
Parametrized field theories are field theories with a background
geometry which however are presented in diffeomorphism covariant form by
promoting the background coordinates to fields. The diffeomorphism
covariance introduces constraints and the physical sector of the theory
is the old theory with a background.  Consider for definiteness a free
field theory on the flat Minkowski space-time and its parametrized form.
In the non-parametrized form, isometries of the Minkowski metric are the
symmetries of the theory. One may choose the usual Fock quantization and
see the representations of the infinitesimal symmetries. In the
parametrized form however, the diffeomorphism covariance would suggest
polymer quantization, not just for the embedding variables but also for
the scalar field.  One can now ask, how the isometries are represented
in such a quantization. 

{\em If} we insist that Dirac quantization of the PFT {\em should}
produce a physical sector which is same as the quantized
non-parametrized theory and it is possible to realise this, then the
quantization of the matter sector chosen in the non-parametrized form
will already determine the symmetry realization, regardless of its
parametrized version. However, it is conceivable that that there is a
(different) Dirac quantization of the polymer quantized PFT such that
the physical states carry the usual Fock representation. There is no
definitive statement available on this as yet\footnote{In two
dimensional space-time, $\mathbb{R}\times S^1$, the work of Laddha and
Madhavan, the third paper of \cite{PFTRecent},  obtains the isometry
group being spontaneously broken to its discrete subgroup.}.  Such a
possibility could be quite relevant for LQG, at least in a
`semiclassical approximation'.  This is one context in which the
discussion of this work, especially the Cyl$^*$ alternative, could be
directly relevant.  }

\acknowledgments 
We would like to thank Alok Laddha and A. P. Balachandran for
discussions.  G.D. would like to thank IUCAA for the warm hospitality
during his visit where this work was concluded. We also thank an
anonymous referee for drawing our attention to parametrized field
theories.

\end{document}